\documentstyle[aps,twocolumn]{revtex}

\begin{document}
\author{G. Gallatin}
\address{SVG Lithography, Wilton, CT}
\title{Mechanism of Laser Induced Compaction}
\date{June 8, 1995}
\maketitle

\begin{abstract}
Fused Silica is the material of choice for UV optical systems such as the
projection optics in microlithography systems. Unfortunately fused silica is
not stable under UV irradition and undergoes compaction and color center
formation. The UV damage properties of fused silica have been the subject of
a number of studies \cite{damage studies}. The detailed microscopic
mechanism for color center formation is complex and remains obscure but the
fact that compaction must be a generic property of a glassy system seems to
have been overlooked by most researchers although Krajnovich, et. al., \cite
{damage studies} do briefly discuss compaction with respect to a specific
glass model. The purpose of this note is to briefly discuss how the basic
mechanism of compaction follows directly from the modern theoretical
understanding of glassy systems \cite{klinger}. This understanding of the
basic mechanism of compaction fits the experimental data quite well. It also
allows for the prediction of some of qualitative properties of compaction
which have not so far been experimentally determined, such as, the absence
of a minimum damage threshold for the onset of compaction and the eventual
long term cessation of compaction after sufficient shrinkage has occurred.
\end{abstract}

Glassy systems such as fused silica consist of random arrangements of atoms
with no long distance order as opposed to the crystalline forms of the same
material which have long distance order. Glassy systems are not at thermal
equilibrium but are stuck in metastable states with Free energies larger
than that of the crystal form at the same temperature\cite{klinger}. This
leads to the density of the glassy system being lower than that of the
crystalline form. In the case of fused silica the density is about 5\% less
than that of the crystalline form, quartz. The exact detailed distribution
of metastable states cannot currently be derived from first principles but
the generic properties of the energy surface can be estimated by comparison
with theoretical results derived for spin glasses$\cite{spin glass}$. These
results indicate that the energy surface is very rough, containing many
local minima or valleys separated by energy barriers which are large
compared to the thermal energy. Also there will be one global minimum
corresponding to the crystalline configuration. The spectrum of energies in
the valleys is almost continuous. Since those valleys corresponding to
amorphous and polycrystalline distributions of atoms are only local minima
of the energy they represent metastable states of the material. Since the
polycrystalline form of the material is ``midway'' between the amorphous and
single crystal forms the energy in polycrystalline valleys is intermediate
between the amorphous and single crystal values. The energy barriers between
microscopically similar states, amorphous or polycrystalline, are low, being
on the order of the bandgap energy. This is because only a few atomic bonds
need to be broken to convert one state into another microscopically similar
state. The energy barriers between macroscopically different forms will be
very high because many bonds need to be broken. Specifically, the energy
barrier between a state containing $N_c$ crystallites and $N_c-1$
crystallites where $N_c$ is much smaller than the number atoms in the sample
(so that the crystallites are macroscopic) is very large compared to the
bandgap. This is because, effectively, all the bonds on the surface of a
given crystallite must be broken in order to macroscopically align it with
respect to an adjacent crystallite thus producing a single larger crytallite
from the original two smaller crystallites. These considerations lead to the
schematic representation of the energy surface shown in Figure 1.

Compaction can be understood in this ``model'' as the following process:
Absorption of photons from the excimer beam provides sufficient energy to
kick the system over the small energy barrier separating one metastable
state from the next lowest metastable state. Successive two photon
absorptions thus gradually allow the material to settle into lower and lower
energy states with higher and higher density. In other words, compaction is
just the material trying to reach its true equilibrium configuration which
is the higher density lower energy crystalline form. The states occupied by
the material will be less and less amorphous and more and more
polycrystalline as the energy of the configuration decreases. As discussed
above, as the mean crystallite size increases so does the energy barrier
separating the current state from one with one less crystallite. Thus based
on this ``model'' we would expect that the rate of compaction should slow
very gradually and eventually cease altogether when the energy barriers
become too high to be surmounted by two photon absorption. Multiphoton
absorption does occur but its rate is orders of magnitude smaller than that
for two photon absorption and hence it has a correspondingly small effect on
compaction.

Two photon absorption is a natural process that does not require ``defects''
or ``impurities'' in the material. Simply put, if the combined energy of two
photons is greater than the bandgap of the material they can be absorbed.
Perfectly pure and defect free material will absorb sufficiently energetic
photon pairs. After all, it is the interaction of the photons with the atoms
of the material that leads to the index of refraction in the first place.
This is not to say that defects and impurities don't play a role at all.
Certainly some absorption processes involve E' centers and/or impurities
like OH radicals and aborption can produce dangling bonds or other defects.
Thus the ``model'' indicates that there should not be any relation between
compaction and color center formation. This is seen experimentally \cite
{damage studies}.

Further, each two photon absorption event is local in nature. That is, the
energy from the two absorbed photons is transferred to a small cluster of
atoms which, following the above ``model'' use it to rearrange themselves
closer to the crystalline form of the material. Thus the energy barriers
between microscopically similar amorphous states must be less than the
energy contained in a single pair of photons which is about 10eV for 248nm
photons. From the above ideas it follows that only two photons at at time
are needed for each single ``compaction step''. Lowering the intensity of
the excimer beam only lowers the number of photons in the beam it does not
change the energy carried by each photon. Thus as the laser fluence is
lowered compaction will continue proportional to the square of the intensity
all the way to essentially zero laser fluence. (NOTE: At 30mJ/pulse there
are on the order of $10^{16}$ 248nm photons in each pulse. So, to get to the
limiting case of just two photons per pulse we would have to lower the
energy by 16 orders of magnitude.) This ``model'' explains the continuity of
compaction as seen in the experiments\cite{damage studies}. That is,
compaction starts immediately as soon as the excimer beam impinges on the
material and continues to increase linearly with the number of pulses
showing no threshold or other sudden transient behavior.

\end{document}